\documentstyle[preprint,aps]{revtex}   

\begin{document} 
\title{Exact expression for the diffusion propagator in a family of    
time-dependent anharmonic potentials}   
\author{ 
J. A. Giampaoli\thanks{Present address: Center for Industrial Research, FUDETEC;  
Av.C\'ordoba 320, 1er. piso; 1054-Buenos Aires; Argentina} 
, D. E. Strier\thanks{Electronic address:strier@df.uba.ar, Present address: Departamento  
de F\'{\i}sica; Facultad de Ciencias Exactas; Universidad de Buenos Aires; Pabell\'on I,  
Ciudad Universitaria, 1428-Buenos Aires; Argentina}, C. Batista, German Drazer 
\thanks{Electronic address:gdrazer@tron.fi.uba.ar, Permanent address:   
Grupo de Medios Porosos, Facultad de Ingenier\'{\i}a, Universidad de Buenos Aires,  
Paseo Col\'on 850. CP 1063, Argentina} and H. S. Wio\thanks{E-mail:~wio@cab.cnea.gov.ar;  
\newline http://www.cab.cnea.gov.ar/CAB/invbasica/FisEstad/estadis.htm}} 
\address{Comisi\'on Nacional de Energ\'{\i}a At\'omica, Centro At\'omico   
Bariloche and Instituto Balseiro (CNEA and UNC); 8400-San Carlos de   
Bariloche, Argentina.}  
\date{\today}  
\maketitle    
\begin{abstract}    
We have obtained the exact expression of the diffusion propagator in the time-dependent anharmonic  
potential $V(x,t)=\frac{1}{2}a(t)x^2+b\ln x$. The underlying Euclidean metric of the problem  
allows us to obtain analytical solutions for a whole family of the elastic parameter $a(t)$,     
exploiting the relation between the path integral representation of the short time propagator and  
the modified Bessel functions. We have also analyzed the conditions for the appearance of a non-zero  
flow of particles through the infinite barrier located at the origin ($b<0$). 
\end{abstract}     
\vspace*{.5cm}    
    
\section{Introduction}    
    
The mathematical theory of stochastic processes has proven to be not only     
a useful but also a necessary tool when studying physical, chemical and     
biological systems under the effect of fluctuations \cite{vkgw}.     
Recent theoretical and experimental studies have shown that there are many  
situations where fluctuations plays an essential role leading to     
new phenomena induced by the presence of noise. A few examples of such     
situations are: some problems related with self-organization and dissipative     
structures \cite{np,dw}, noise induced transitions \cite{hl},     
noise induced {\it phase} transitions \cite{nipt}, noise sustained     
patterns \cite{nsp}, stochastic resonance in zero-dimensional      
and in spatially extended systems \cite{SR1,SR2}.     
    
An almost natural way to describe (Markovian) stochastic processes     
corresponds to an approach introduced by Wiener  based on a     
sum over trajectories \cite{wie}, anticipating by two decades Feynman's      
work on path integrals \cite{fey}. This approach was later applied by Onsager and     
Machlup to some Markovian nonequilibrium processes \cite{om}.     
However, even some non-Markov processes can be also     
described within this framework \cite{col}. In spite of the historical     
fact, path and functional integration methods have been largely studied     
and applied within the quantum mechanical realm, while its study and     
application in stochastic processes is scarce \cite{qmsp}.     
However, it is worth here to remark that, following earlier results \cite{critic1},  
path and functional integral methods have been widely applied in the theory of  
dynamic critical phenomena, both at equilibrium and nonequilibrium phase transitions  
\cite{critic2}.  
 
Among others, one of the problems that has not received much attention  
within the path integral description of stochastic processes is the application     
of space-time  transformations, while this kind of transformations     
have been largely used within the realm of quantum mechanics \cite{tsp}.     
Among the few studies in this regard (see Ref. \cite{old}) a recent one refers to      
a transformation relating the diffusion propagator in a     
time-dependent {\it harmonic} oscillator     
with the propagator for the case of free diffusion, including a whole     
family of possible analytical solutions \cite{bat}.     
Also, a formal adaptation of Duru-Kleinert-like transformations to the     
stochastic case, overcoming the main disadvantage of the direct     
application of such transformations --namely that Duru-Kleinert     
transformations do not link different Markov processes-- was     
introduced in Ref. \cite{kle2}. However, it merits to stand out     
the treatment of transformations between ``different" Wiener     
processes done in Friendlin's book (indicated with (j) in     
Ref. \cite{qmsp}), even though they are not the exact equivalent to     
the Duru-Kleinert transformations.     
    
In this paper we present the exact solution for the diffusion propagator     
in a time-dependent anharmonic oscillator $V(x,t)=\frac 12a(t) x^2+b\ln x$.  
This particular choice of the potential can be useful to model the behavior  
of several physical and biological systems. Among them, the study of neuron models  
(e.g. integrate and fire models \cite{SR1}),  
stochastic resonance in monostable nonlinear 
oscillators \cite{SR3} and its possible application to  
spatially extended systems \cite{SR2}.  
Also, we can consider that the logarithmic term, whithin the  
path integral scheme (or as a Boltzmann like weight) mimics a prefactor corresponding  
to an effective energy barrier.  
It is clear that the possibility of having  
exact expressions of the stochastic propagator in a non-symmetrical potential can be  
of interest. In fact, in many of the above mentioned applications (specially on neuron  
models) the potential studied in this work would represent a more realistic approximation  
to the real behavior of the system under study. In other problems such us brownian motors,  
this asymmetry is not just an improvement but an unavoidable ingredient of the model. 
 
The approach used in this work has been inspired by the solution presented  
in Ref.\cite{kl}, corresponding to the exact quantum mechanical propagator 
in a  
time-dependent harmonic potential plus a singular perturbation.  
In the present case, the fact that the metric of the underlying space is 
Euclidean, allows us to obtain the exact analytical expression of the 
diffusion propagator for a whole family of functional forms of the 
time-dependent elastic parameter. 
   
In the next section we introduce the model we are going to study and show     
the procedure to be followed in order to obtain the exact form of the     
propagator. We also discuss the presence of noise induced flow of particles 
through the infinite barrier located at the origin, provided that the noise  
amplitude is large enough for the particles to overcome the deterministic drift. 
In section \ref{analytical_solutions} we show how to obtain     
a family of analytical solutions. In the     
last section we make a final discussion and comment on the possible     
applications of the present results.     
    
\section{The diffusion propagator}    
\label{propagator}    
In this section we will follow, and adequately adapt, the results of the     
paper of Khandekar and Lawande \cite{kl}.     
Our starting point is to consider the following Langevin equation    
 \begin{equation}    
\dot{x}=h(x,t)+\xi (t),       
\label{lang}    
\end{equation}    
where $\xi (t)$ is an additive {\it Gaussian white noise} \cite{vkgw}.     
That is it fulfills the conditions $\langle \xi (t) \rangle= 0$ and     
$\langle \xi (t) \xi (t') \rangle = 2 D \delta (t-t')$.     
This equation describes the overdamped motion of a particle in a     
time-dependent potential. In this work, we will consider the force term     
$h(x,t)=-a(t)x-b/x$ which, through the relation      
$h(x,t)=-\partial V/\partial x$, corresponds to the following potential    
\begin{equation}    
V(x,t)=\frac 12a(t)x^2+b\ln x.      
\label{pot}    
\end{equation}                   
This potential is defined for $x>0$ and, whenever $b<0$ and the elastic  
parameter $a(t)$ is positive, it corresponds to an anharmonic monostable system  
composed by a time-dependent harmonic oscillator plus a logarithmic term  
which is singular at the origin.  
 
As it will be shown later, even in this monostable situation,  
the noise could be able to induce a flow of particles through the infinite    
barrier located at the origin, overcoming the deterministic drift whenever    
$D>-2b$ holds. In fact, the meaningful condition related to the    
conservation of particles inside the system (zero flux at $x=0$) is $D<-2b$.   
 
In this work we will relax the monostability condition,  
allowing for the time-dependent elastic parameter to take  
negative values. We will show that in this extended situation an  
asymptotic probability distribution can be reached whenever the  
elastic term satisfies, 
\begin{equation}\label{asymptotic} 
\lim_{t_0 \to -\infty}\int_{t_0}^t a(s)ds = \infty \qquad \forall \, t. 
\end{equation} 
We will say that in this case the potential is {\it strongly attractive}.     
 
The  path integral representation of $P(x_b,t_b \mid x_a,t_a)$, that is the     
transition probability associated with this Langevin equation, is given     
by  (see for instance the book indicated as "i" in \cite{qmsp})    
\begin{equation}    
\label{ptr1}     
P(x_b,t_b \mid x_a,t_a)=\int_{x(t_a)=x_a}^{x(t_b)=x_b} {\cal D}[x(t)]    
\;\; \exp \left[ -\int_{t_a}^{t_b}L(x(\tau ),\dot{x}(\tau ),\tau    
)d\tau \right].     
\end{equation}    
Here the stochastic {\it Lagrangian }or {\it Onsager-Machlup} \cite{om}    
functional is given, in a midpoint discretization, by      
\begin{equation}    
\label{lag1}    
L(x,\dot{x},t)=\frac 1{2D}[\dot{x}-h(x,t)]^2+\frac 12\frac{%
\partial h(x,t)}{\partial x}.     
\end{equation}    
Replacing the actual form of $h(x,t)$ the previous expression can be    
expanded to yield    
\begin{eqnarray}    
L &=& L_0+\frac{d\Phi }{dt}, \label{lag2}     
\end{eqnarray}    
where $\Phi $ corresponds to     
\begin{equation}    
\Phi (t)=\left[ \frac bD-\frac 12\right] \int_{t_{0}}^t a(\tau    
)d\tau +\frac bD\ln x+\frac{a(t)x^2}{2D},  \label{phi}    
\end{equation}    
with arbitrary $t_0$, and    
\begin{equation}    
L_0\equiv \frac 1{2D}\left[ \dot{x}^2+\left( a(t)^2-\dot{a}%
(t)\right) x^2+\left( b+D\right) \frac b{x^2}\right].  \label{lag0}    
\end{equation}    
    
Hence the path integral in Eq. (\ref{ptr1}) adopts the form    
\begin{equation}    
P(x_b,t_b\mid x_a,t_a)=e^{-[\Phi (t_b)-\Phi (t_a)]}K(x_b,t_b\mid x_a,t_a),     
\label{ptr2}    
\end{equation}    
with    
\begin{equation}    
\label{ker1}     
K(x_b,t_b \mid x_a,t_a)=\int_{x_a}^{x_b}{\cal D}[x(t)]    
\;\; \exp \left[ -\frac 1{2D}\int_{t_a}^{t_b}\left( \dot{x}%
^2+\omega (\tau )x^2+\left( b+D\right) \frac b{x^2}\right) d\tau \right], \\    
\end{equation}    
and $\omega (t)=a(t)^2-\dot{a}(t).$ As usual, the path integral in     
Eq. (\ref{ker1}) is defined  in a discretized form by    
\begin{equation}    
K(x_b,t_b\mid x_a,t_a)=\lim_{N\rightarrow \infty }A_N\int ...\int \exp    
\left( -\sum_{j=1}^N S_j(x_j,x_{j-1})\right) \prod_{j=1}^{N-1}dx_j    
\label{ker2},      
\end{equation}    
with $N \varepsilon =t_a - t_b$, $t_j= t_a+ j \varepsilon $,     
$x_0 =x_a$; $x_N=x_b$; $A_N=\left[{2\pi D\varepsilon }\right]^{-\frac{N}{2}}$;     
and     
\begin{eqnarray*}    
S_j=S_j(x_j,x_{j-1}) &=&\varepsilon \;L_o(x_j,x_{j-1})\;=\frac 1{2D}\left[ \frac{%
x_j^2+x_{j-1}^2}\varepsilon +\varepsilon \omega _jx^2\right] -\left[ \frac{%
x_jx_{j-1}}{D\varepsilon }+\frac{D(\theta ^2-\frac 14)\varepsilon }{%
2x_jx_{j-1}}\right].     
\end{eqnarray*}    
Here    
\begin{equation}    
\theta = \theta (b) = \frac{1}{2}\sqrt{1+\frac{4b(b+D)}{D^2}} = \left| \frac{1}{2} + \frac{b}{D} \right| .    
\label{tita}    
\end{equation}    
    
Up to first order in $\varepsilon $ (exploiting that $\varepsilon \ll 1$)    
we can use the following asymptotic form of the modified Bessel function    
\begin{equation}    
\exp \left( \frac{ u}{\varepsilon} -\frac{1}{2}\left( \theta (b)^2-\frac 14\right)    
\frac{\varepsilon }{u}+ O(\varepsilon ^2)\right) \approx \sqrt{\frac{%
2\pi u}\varepsilon }\;I_{\theta (b)}\left( \frac u\varepsilon \right).   
\label{ide}    
\end{equation}    
 
Using the last expression with $u=x_j\,x_{j-1}\,/\,D$ the propagator of  
Eq.(\ref{ker2}) may be cast to the following form    
\begin{eqnarray}    
K(x_b,t_b \mid x_a,t_a)&=&\lim_{N\rightarrow \infty }\left( \frac 1{2\pi    
D\varepsilon }\right) ^{1/2}\int ...\int \prod_{j=1}^{N-1}dx_j    
\label{ker3} \\    
&&\times \prod_{j=1}^N \exp \left[ -\frac    
1{2D\varepsilon }(x_j^2+x_{j-1}^2+\varepsilon ^2\omega _jx_j^2)\right]     
\left( \frac{2\pi x_jx_{j-1}}{D\varepsilon }\right) ^{1/2}I_\theta    
\left( \frac{x_jx_{j-1}}{D\varepsilon }\right).     
\nonumber    
\end{eqnarray}    
The last expression can be rewritten as    
\begin{equation}    
\label{ker4}     
K(x_b,t_b \mid x_a,t_a)=\exp \left[ \frac{-\beta }2\left(    
x_a^2+x_b^2\right) \right] \lim_{N\rightarrow \infty }\beta ^N\;\;    
\int ...\int \prod_{j=1}^{N-1}e^{-\alpha _j\;x_j^2}\;I_\theta \left( \beta    
x_jx_{j-1}\right) x_jdx_j,      
\end{equation}    
where,    
\[    
\alpha _j=\beta \left( 1+\frac{\varepsilon ^2}2\omega _j\right); \;\;\;0\leq    
j\leq N-1;\;\;\text{and}\;\;\beta =\frac 1{D\varepsilon }.      
\]    
Now, in order to perform the integrations of Eq.(\ref{ker4}), we can use     
the equality known as Weber formula \cite{wat,pi}, which is given by     
\begin{eqnarray*}    
\int_0^\infty e^{-\alpha x^2}\;I_\theta (ax)\;I_\theta (bx)\;x\;dx &=&\frac    
1{2\alpha }\exp \left[ \frac{a^2+b^2}{4\alpha }\right] I_\theta \left( \frac{%
ab}{2\alpha }\right) ,     
\end{eqnarray*}    
and is valid for     
${\Re}\left( \theta \right) >-1,\;\;{\Re}(\alpha )>0$     
(here both conditions are fulfilled). The final result is     
\begin{equation}    
K(x_b,t_b\mid x_a,t_a)=\sqrt{x_ax_b}\lim_{N\rightarrow \infty    
}a_N\;e^{\left( p_Nx_a^2+q_Nx_b^2\right) }I_\theta \left( a_Nx_ax_b\right),      
\nonumber    
\end{equation}    
where the quantities $a_N$, $p_N$ and $q_N$ are defined in appendix A.     
These quantities are related to a function $Q(t)$ that obeys the equation     
(as usual, we have indicated time derivatives with dots)     
\begin{equation}    
\ddot{Q}(t)-\omega (t)Q(t)=0,   \label{mec}    
\end{equation}    
with the initial condition $\;\;Q_0=Q(t_a)=0 $.     
In the limit $\varepsilon \rightarrow 0\;(N\rightarrow \infty )$,     
we find that (see appendix A)    
\begin{equation}    
\lim_{N\rightarrow \infty }a_N=\frac 1D\frac{\dot{Q}(t_a)}{Q(t_b)}, \label{liman} 
\end{equation}    
\begin{equation}    
\lim_{N\rightarrow \infty }p_N=\lim_{\varepsilon \rightarrow 0}\left( \frac    
1\varepsilon -\dot{Q^2}(t_a)\int_{t_a+\varepsilon}^{t_b}    
\frac{dt}{Q(t)^2}\right),  \label{limpn}  
\end{equation}    
\begin{equation}    
\lim_{N\rightarrow \infty }q_N=-\frac 1{2D}\frac{\dot{Q}(t_b)}{Q(t_b)}, \label{limqn}.  
\end{equation}    
To calculate the second limit, it is necessary to solve Eq. (\ref{mec}).     
As it was shown in Ref. \cite{bat}, the complete solution of     
Eq. (\ref{mec}) can be reduced to {\it quadratures}, with the general     
form given by     
\begin{equation}\label{gsol1}     
Q(t)=k_1\;{\cal R}_{t_a}(t)+k_2\;{\cal S}_{t_a}(t),     
\end{equation}    
 
\noindent where 
\begin{eqnarray} 
{\cal R}_{t_a}(t)&=&e^{-\int_{t_a}^ta(s)ds} \\ \nonumber 
{\cal S}_{t_a}(t)&=&e^{-\int_{t_a}^t a(s) ds} \int_{t_a}^{t}e^{2\int_{t_a}^{\tau} 
 a(\varsigma )d\varsigma } d\tau.  \label{gsol2} 
\end{eqnarray} 
   
Hence, the solution fulfilling the initial condition $Q_0=Q(t_a)=0$ is    
\begin{equation}    
Q(t)=\dot{Q}(t_a)\;{\cal S}_{t_a}(t). \label{defs}    
\end{equation}    
 
After replacing this solution into the expressions for $a_N$, $p_N$     
and $q_N$ (see appendix B), we finally arrive to a completely analytical     
expression for the transition probability    
\begin{eqnarray}    
P(x_b,t_b \mid x_a,t_a)\!\!&=&\!\!e^{-[\Phi (t_b)-\Phi (t_a)]}\frac{\sqrt{x_ax_b}} 
{D\;{\cal S}_{t_a}(t_b)}\times I_\theta \left(\frac{x_ax_b}{D\;{\cal S}_{t_a}(t_b)}\right) \times \label{trpold}    
\\    
&& \hspace{-1cm} \exp \left( \frac{-1}{2D\;{\cal S}_{t_a}(t_b)}\left[ \Bigg( {\cal R}_{t_a}(t_b)+ 
a(t_a)\Bigg) x_a^2 +     
\left(\frac{1}{{\cal R}_{t_a}(t_b)}-a(t_b)\right) x_b^2\right]\right),  \nonumber 
\\ \nonumber 
\end{eqnarray}    
which can be further simplified as    
\begin{eqnarray}    
P(x_b,t_b \mid x_a,t_a)\!\!&=&\!\!x_a^{\frac{b}{D}+\frac{1}{2}}x_b^{-\frac{b}{D}+\frac{1}{2}}\frac{[{\cal R}_{t_a}(t_b)]^{\frac{b}{D}-\frac{1}{2}}}{D\;{\cal S}_{t_a}(t_b)}\times I_\theta \left(\frac{x_ax_b}{D\;{\cal S}_{t_a}(t_b)}\right) \times \label{trp}    
\\    
&& \hspace{-1cm} \exp \left( \frac{-1}{2D\;{\cal S}_{t_a}(t_b)}\left[{\cal R}_{t_a}(t_b) x_a^2 +     
\frac{1}{{\cal R}_{t_a}(t_b)} x_b^2\right]\right).  \nonumber 
\\ \nonumber 
\end{eqnarray}    
 
It is straightforward to check for some particular choices of $a(t)$ and $b$ that the last expression  
fulfills the corresponding Fokker-Planck equation. Albeit not so simple, we have also proved it for  
the general case. 
The last expression also indicates that, in order to have the explicit form     
of the propagator we only need to obtain the     
function $Q(t)$ (the solution of Eq. (\ref{mec}) given in Eq. (\ref{gsol1}))    
for the problem under study (that is, for a given form of the function     
$a(t)$). We will provide a family of solutions for a rather general form of the function  
$a(t)$ in a subsequent section. Before, we will discuss the possibility of finding a net current at the  
origin.  
 
\subsection{Flow through the infinite barrier} 
\label{zero_flow}    
Let us first evaluate from Eq. (\ref{trp}) the asymptotic probability  
distribution, that is  
\begin{equation} 
P(x,t)=\lim_{t_a\to-\infty}P(x,t \mid x_a,t_a).  
\end{equation} 
 
In the strongly attractive  case (Eq. (\ref{asymptotic})), it can be easily  
shown that  
${\cal S}_{t_a}(t)$ diverges and that ${\cal R}_{t_a}(t)$  
goes to zero  as $t_a \to - \infty$.  
Thus, we will make use of the expansion of the modified Bessel function for small argument \cite{wat}, 
\begin{equation} 
I_{\theta}(z) = \frac{1}{\Gamma(\theta +1)}\, \left( \frac{z}{2}\right)^{\theta} + {\cal O}(z^{\theta+2}).  
\end{equation} 
Replacing this expansion into Eq. (\ref{trp}) we get, 
\begin{equation}    
\label{asym1} 
P(x,t)\!\!=\!\!\;\left[ \lim_{t_a \to -\infty} 
[x_a \, {\cal R}_{t_a}(t)]^{\frac{b}{D}+\frac{1}{2}+\theta}\right] 
\frac{2}{\Gamma(\theta+1)} 
\frac{x^{-\frac{b}{D}+\frac{1}{2}+\theta}}{(2\;D\;g(t))^{1+\theta}} 
\exp \left(-\frac{x^2}{2\;D\;g(t)}\right),  \nonumber 
\end{equation}    
where $g(t)$ is defined as 
\begin{equation} 
g(t) = \lim_{t_a \to -\infty} {\cal S}_{t_a}(t) {\cal R}_{t_a}(t).  
\label{g} 
\end{equation} 
It is clear that unless the condition  
\begin{equation}\label{condition} 
1/2+b/D+\theta=0  
\end{equation} 
holds, the system cannot reach an asymptotic probability distribution. In fact,  
the term between the square brackets in Eq. (\ref{asym1}) depends on the initial condition.  
Furthermore, it can be shown that the normalization of Eq. (\ref{asym1}) gives a vanishing  
function of $t$ unless the previous condition holds. 
Note that Eq. (\ref{condition}) implies, through Eq. (\ref{tita}), $D<-2b$. This condition gives  
the maximum value of noise amplitude for the particles to be confined inside the interval  
$(0,\infty]$. 
 
This encourage us to show explicitly the existence of a noise induced probability current  
through the infinite barrier at the origin when  $D>-2b$. Before giving a  
rigorous deduction, let us state a simple argument which provides some clue  
about the underlying physical mechanism governing this flow. As it is clear  
from the Langevin equation, the particle is subjected to both deterministic and stochastic forces.  
If we analyze separately both contributions to the particle movement near the origin, we obtain  
for the deterministic trajectory $x_{d}(t)=\sqrt{-2bt}$. Comparing this result with the well  
known diffusive behavior, where the uncertainty on the particle's position grows as  
$x_{s}=\sqrt{Dt}$, we reobtain the previous condition $D>-2b$ for the possible appearance of  
noise induced leakage of particles.  
 
The probability current at the origin $J(x=0,t\mid x_a,t_a)$ can be evaluated from the  
associated Fokker-Planck equation. In the case $D>-2b$ we obtain:  
\begin{equation} 
J(x=0,t\mid x_a,t_a)=\left.\left(-ax-\frac{b}{x}-\frac{D}{2}\frac{\partial}{\partial x}\right) 
P(x,t\mid x_a,t_a)\;\right|_{x=0}=-D\;\left(\frac{b}{D}+\frac{1}{2}\right)J_{x_a,t_a}(t),  
\end{equation} 
where it can be shown that $J_{x_a,t_a}(t)$ is a positive function of time for any given  
initial condition. Therefore, we have obtained a non-zero negative current, as previously  
stated\footnote{This somewhat counter-intuitive flux has an interesting  
quantum counterpart in the ``fall to the center'' effect studied by Landau \cite{landau}.}.    
 
\subsection{Asymptotic probability distribution} 
 
In the case $D<-2b$ there is no probability leakage. In fact, the asymptotic probability  
distribution can be obtained from Eq. (\ref{asym2}) and is given by: 
\begin{equation} 
\label{asym2} 
P(x,t)\!\!=\!\!\; 
\frac{2}{\Gamma(\theta+1)} 
\frac{x^{-\frac{2b}{D}}}{(2\;D\;g(t))^{1+\theta}} 
\exp \left(-\frac{x^2}{2\;D\;g(t)}\right) 
\end{equation} 
which can be easily shown to be normalizable. 
 
It is worth to study how the properties of the elastic parameter 
function $a(t)$ influences the behavior of the function $g(t)$, which 
reflects the time evolution of the width of the probability distribution.  
First, note that from the definition of $g(t)$ given in Eq. (\ref{g}) we 
can obtain  
\begin{equation} 
\dot g (t)=-2a(t)g(t)+1. 
\label{dotg} 
\end{equation}  
From this equation it can be deduced that $g(t)>0$ $\forall t$, as 
must be expected for any well behaved probability distribution. 
In addition, it can be proved that in order to confine the particle 
in a small region of width  $\sqrt{g(t)} \sim \sqrt{\epsilon}$ an 
attractive force of order $a(t)\sim 1/\epsilon$ is needed.  
On the other hand, a small attractive force of order $a(t)\sim \epsilon$, 
gives a broad distribution  
with $g(t)\sim 1/\epsilon$. The limiting cases for $P(x,t)$ corresponding to 
an unbounded  spreading ($a(t) \to 0 \Rightarrow g(t) \to \infty$), and to 
an asymptotically approach to a $\delta(x)$ distribution  
($a(t) \to \infty \Rightarrow g(t) \to 0$), can be also obtained.  
 
From the previous paragraph it is clear that even in the strongly 
attractive situation  
(see Eq. (\ref{asymptotic})) the probability distribution may exhibit 
an unbounded spreading.  
In fact, we have already shown that even in the monostable situation 
$a(t)>0$, but where the  
strength vanishes in time,  $g(t)$ grows indefinitely. Therefore, in order 
to obtain a  
non-divergent  width of the probability distribution, the conditions 
on the attractive  
term have to be stronger than the one imposed by Eq. (\ref{asymptotic}). 
We may infer that the localized-probability condition should be related to a  
non-vanishing attractive strength of the time averaged potential.  
 
\subsection{General localization conditions} 
 
Let us discuss the set of conditions that ensures the asymptotic localization  
of the probability distribution.  
From the analysis of Eq. (\ref{dotg}) it is clear that in order to guarantee  
a non divergent $g(t)$ the elastic parameter should have the following 
properties.  
Its accumulated strength is positive, i. e. 
\begin{equation} 
\int_{t_i}^{t}a(\tau) \, d\tau = c > 0  \qquad \forall t   
\label{c.a.c.a.} 
\end{equation} 
where $c$ is an arbitrary constant and $t_i$ is the nearest time which 
fulfills the previous equation.  
This condition is clearly fulfilled if the potential is  
strongly attractive. We may also infer that the accumulated attractive 
effect (Eq. (\ref{c.a.c.a.}))   
should be non-vanishing. In other words, the elapsed time where the 
accumulated strength reaches  
the given constant $c$ is bounded, that is    
\begin{equation} 
t-t_i = \Delta t \leq \Delta t_{u} \qquad \forall t 
\label{deltat} 
\end{equation} 
where $\Delta t_{u}\equiv\Delta t_u (c)$ is the mentioned upper bound for the elapsed time.  
It is evident that condition (\ref{deltat}) is more restrictive than the one imposed by Eq. (\ref{asymptotic}). 
 
In the following section we will provide a family of examples where the 
probability is asymptotically localized.  
 
\section{A family of analytical solutions}    
\label{analytical_solutions}    
As already mentioned, to obtain the final expression for the diffusion propagator, we must first  
solve Eq. (\ref{mec}) for a given choice of the function $a(t)$. Because the frequency  
$\omega (t)$ depends only on the harmonic term of the potential, we can make use of any known solution 
of the simpler harmonic case. A method to generate a whole family of analytical solutions has been  
proposed in Ref.\cite{bat} for the time-dependent harmonic oscillator. In     
order to reach such a goal the elastic parameter was written in the     
following form    
\begin{equation}    
a(t)=f(t)+\frac 12\frac{\dot{f}(t)}{f(t)}  \label{fat}    
\end{equation}    
This allows us to find the corresponding independent solutions of Eq. (\ref{mec})    
\begin{equation}    
q_1(t)=\frac{\sinh (F(t))}{\sqrt{f(t)}}  \label{q1t}    
\end{equation}    
\begin{equation}    
q_2(t)=\frac{\cosh (F(t))}{\sqrt{f(t)}},  \label{q2t}    
\end{equation}    
where $F(t)=\int_{t_0}^t f(s)\;ds$, indicating that $f(t)$ must be an     
integrable function.    
The solution which satisfies the initial condition $Q(t_a)=0$   
reduces to     
\begin{equation}    
Q(t)=\dot{Q}(t_a)\frac{\sinh (F(t)-F(t_a))}{\sqrt{f(t)f(t_a)}}.      
\label{defs2}    
\end{equation}    
With this result, the transition probability in Eq. (\ref{trp}) adopts the     
analytical form    
\begin{eqnarray}    
P(x_b,t_b &\mid &x_a,t_a)=  {\rm e}^{\left[\frac{1}{2}-\frac{b}{D}\right](F(t_b)-F(t_a))} 
\left(\frac{f(t_a)}{f(t_b)}\right)^{\frac{b}{2D}-\frac{1}{4}} 
x_a^{\frac{b}{D}+\frac{1}{2}} x_b^{-\frac{b}{D}+\frac{1}{2}} 
\frac{\sqrt{f(t_b)f(t_a)}}{D\sinh (F(t_b)-F(t_a))}\nonumber \\ 
&&\times I_{\theta (b)} \left( \frac{x_ax_b}D%
\frac{\sqrt{f(t_b)f(t_a)}}{\sinh (F(t_b)-F(t_a))}\right) \nonumber \\    
&& \exp \left( -\frac{ f(t_a)%
e^{-(F(t_b)-F(t_a))} x_a^2 + f(t_b) e^{(F(t_b)-F(t_a))} x_b^2}{2D\sinh (F(t_b)-F(t_a))} \right).      
\label{trp2}    
\end{eqnarray}    
Hence, we have obtained a completely analytical expression for the      
propagator in Eq. (\ref{trp2}), that only depends on the choice of     
the elastic parameter $a(t)$.     
    
\section{Final Remarks}    
\label{remarks} 
In this work we obtained the exact expression for the diffusion  
propagator in the time dependent anharmonic potential  
$V(x,t)=\frac 12a(t) x^2+b\ln x$ for a rather general choice of the elastic parameter. 
The knowledge of the exact form of the propagator can be useful to model 
different physical and biological phenomena. Particularly interesting problems, 
suitable to be studied taking advantage of this results, are 
realistic non-symmetric neuron membrane potentials \cite{further} and the phenomenon 
of stochastic resonance in  a monostable zero-dimensional potential,   
in spatially extended systems, and in several neuron firing models, among others.  
A complete and recent review of these stochastic resonance topics can be found in  
the work of Gammaitoni {\it et al}\cite{SR1}. 
On the other hand, the knowledge of the exact propagator  
in the indicated time-dependent anharmonic potential can be useful as a benchmark to test     
approximate numerical or analytical procedures. Among them we can refer to some 
of the problems discussed in \cite{lutz}.  
 
Among the several studies of stochastic resonance in monostable systems, it has been shown using  
scaling arguments and numerical experiments, that the signal to noise ratio is a monotonically  
increasing function of the noise amplitude \cite{mono}. By contrast,   
it is quite clear that this increase in the response of the system cannot be unbounded. 
We shall present elsewhere our own results concerning the phenomenon of stochastic resonance in a  
system described by the potential discussed in the present paper. 
In this regard, the relation between the maximum noise amplitude and the deterministic  
force near the origin proved to be a meaningful cutoff for the increase of the response. 
 
It is worth to remark here that the limit $b \to 0$ is a 
(kind of) singular one. The naive point of view will be that, in 
such a limit, the form of the propagator in Eq. (\ref{trp}) shall 
reduce to the one     
corresponding to the case of the harmonic time-dependent potential     
$V(x) \sim \frac 12 a(t) x^2$.   
However, this limit corresponds to  
a harmonic time-dependent potential for $x>0$ with  
an absorbing boundary condition at $x=0$. Then, it should be possible 
to re-obtain the limit $b \to 0$ of the diffusion propagator found 
in this paper ($P_{0}(x_b,t_b \mid x_a,t_a)$) from the one obtained 
in Ref. \cite{bat} for the harmonic time-dependent case 
($P_h(x_b,t_b \mid x_a,t_a)$) simply as $P_{0}(x_b,t_b \mid x_a,t_a)
= P_h(x_b,t_b \mid x_a,t_a)-P_h(-x_b,t_b \mid x_a,t_a)$.
It can be easily proved that this is indeed the case.

\acknowledgements    
DES and HSW thanks for the kind hospitality extended to them during     
their stay at the ICTP, Trieste, Italy. DES thanks for support from UBA     
through a FOMEC grant, Argentina, and HSW for partial support from CONICET,     
and ANPCyT (Argentinian agencies) and from CEB, Bariloche, Argentina.

\appendix    
    
\section*{A}    
    
The quantities $a_N$, $p_N$ and $q_N $ are defined according to     
    
\[    
a_N=\beta \prod_{j=1}^{N-1}\frac \beta {2\gamma _j};\;\;\;\;\gamma _1=\alpha    
_1,\;\;\;\;\gamma _j=\alpha _j-\frac{\beta ^2}{4\gamma _{j-1}}    
\]    
    
\[    
p_{N=}-\frac \beta 2+\sum_{j=1}^{N-1}\frac{\beta _j^2}{4\gamma _{j}}%
;\;\;\;\;\;q_N=-\frac \beta 2+\frac{\beta ^2}{4\gamma _{N-1}};\;     
\]    
    
\begin{equation}    
\;\beta _1=\beta ;\;\;\;\;\;\beta _j=\beta \prod_{k=1}^{j-1}\frac \beta    
{2\gamma _k},     
\label{coef}    
\end{equation}    
In order to determine the limiting (when $N \to \infty$) values of     
$a_N$,\ $p_N$,\ and $q_N$, it is usefull to define the following     
auxiliary quantities     
    
\begin{equation}    
\lambda _j=\frac 2\beta \gamma _j\;;\;\;\Lambda _k=\prod_{j=1}^k\frac    
1{\lambda _j},  \label{coef2}    
\end{equation}    
with $\alpha _j$ as defined after Eq. (\ref{ker3}) and $\gamma _j$ adopting the     
form    
    
\begin{equation}    
\gamma_j=\beta \left( 1+\frac{\varepsilon ^2\omega _j}2\right) -\frac{\beta ^2}{%
4\gamma _{j-1}},  \label{alfa}    
\end{equation}    
that allows us to obtain the following equation for $\lambda _j$    
    
\begin{equation}    
\lambda _j=2\left( 1+\frac{\varepsilon ^2\omega _j}2\right) -\frac 1{\lambda    
_{j-1}}.  \label{lam}    
\end{equation}    
If we now define that $\lambda _j=\frac{Q_{j+1}}{Q_j}$, the last equation can be    
rewritten as     
\begin{equation}    
Q_{j+1}-2Q_j+Q_{j-1}=\omega _j\varepsilon ^2Q_j,  \label{qeq}    
\end{equation}    
which, in the limit $N \to \infty$ (and $\varepsilon \to 0$), becomes    
Eq. (\ref{mec}), with the initial condition $\;\;Q_0=Q(t_a)=0 $, that     
follows from Eq. (\ref{qeq}).    
    
Finally, we can express the coefficients $a_N,\;p_N,\;$ and $q_N$ as     
functions of the new variables     
    
\begin{equation}    
a_N=\beta \Lambda _{N-1}=\beta \frac{Q_1}{Q_N}=\beta \frac{Q_1-Q_0}{Q_N}%
=\beta \varepsilon \left( \frac{Q_1-Q_0}\varepsilon \right) \frac 1{Q_N}    
\label{an}    
\end{equation}    
    
\begin{equation}    
p_N=-\frac \beta 2\left( 1-\sum_{j=1}^{N-1}\frac{Q_1^2}{Q_{j+1}Q_j}%
\right)  \label{pn}    
\end{equation}    
\begin{equation}    
q_N=-\frac{\beta \varepsilon }2\left( \frac{Q_N-Q_{N-1}}{\varepsilon Q_N}%
\right).  \label{qn}    
\end{equation}    
    
\section*{B}    
    
The replacement of the general solution for $Q(t)$ indicated in     
Eq. (\ref{defs}) into Eqs. (\ref{liman},\ref{limpn},\ref{limqn}),     
leads us to obtain the limiting values of $a_N$, $p_N$ and $q_N$.     
For $p_N$ we find    
    
\begin{equation}    
\lim_{N\rightarrow \infty }p_N=\lim_{\varepsilon \rightarrow 0}\frac{-1}{2D}%
\left( \frac 1\varepsilon +\frac 1{\int_{t_a}^{t_b}d\tau \;e^{2\int_{t_a}^\tau    
a(\varsigma )d\varsigma }}-\frac 1{\int_{t_a}^{t_a+\varepsilon }d\tau \;    
e^{2\int_{t_a}^\tau a(\varsigma )d\varsigma }}\right).  \label{limi}    
\end{equation}    
    
Making a Taylor expansion up to second order in $\varepsilon $ of the last    
denominator we can calculate the limit in Eq. (\ref{limi}) yielding     
    
\begin{equation}    
\lim_{N\rightarrow \infty }p_N=\frac{-1}{2D}\left( \frac    
1{\int_{t_a}^{t_b}d\tau \; e^{2\int_{t_a}^\tau a(\varsigma )d\varsigma    
}}+a(t_a)\right).  \label{pndef}    
\end{equation}    
The expressions for $a_N$ and $q_N$, in terms of the explicit form for     
$Q(t)$ results in     
    
\begin{equation}    
\lim_{N\rightarrow \infty }q_N=\frac{-1}{2D}\left( \frac{e^{%
\int_{t_a}^{t_b}a(\varsigma )d\varsigma }}{\int_{t_a}^{t_b}d\tau \;    
e^{2\int_{t_a}^\tau a(\varsigma )d\varsigma }}-a(t_b)\right)  \label{defan}    
\end{equation}    
    
\begin{equation}    
\lim_{N\rightarrow \infty }a_N=\frac 1D\frac{e^{\int_{t_a}^{t_b}a(\varsigma    
)d\varsigma }}{\int_{t_a}^{t_b}d\tau \; e^{2\int_{t_a}^\tau a(\varsigma    
)d\varsigma }}.  \label{defqn}    
\end{equation}

\end{document}